\documentclass[12pt]{article}
\usepackage{amsmath,amssymb,bm,epsfig}
\usepackage{hyperref}
\usepackage{graphicx}
\usepackage{float}
\usepackage{subfloat}
\advance \topmargin by -\headheight
\advance \topmargin by -\headsep
     
\evensidemargin \oddsidemargin
\makeatletter
\renewcommand\section{\@startsection {section}{1}{\z@}%
                                   {-3.5ex \@plus -1ex \@minus -.2ex}%
                                   {2.3ex \@plus.2ex}%
                                   {\normalfont\large\bfseries}}
 
\renewcommand\subsection{\@startsection{subsection}{2}{\z@}%
                                   {-3.25ex\@plus -1ex \@minus -.2ex}%
                                   {1.5ex \@plus .2ex}%
       {\normalfont\normalsize\bfseries}}
\renewcommand{\subsubsection}{\@startsection{subsubsection}{3}{0mm}
   {0.75\baselineskip}%
   {0.5\baselineskip}%
   {\normalfont\normalsize\slshape\underline}}%
\makeatother
\renewcommand\d{\partial}
\newcommand\x{\mathbf{x}}

\newcommand\dlr{\raisebox{0.1em}{$\stackrel{\scriptstyle\leftrightarrow}\partial$}}
\newcommand\+{\dagger}

\begin{document}
\title{\bf{On the question of symmetries in non-relativistic diffeomorphism invariant theories }}


\author{
{\bf {\normalsize Rabin Banerjee}$^{a}
$\thanks{rabin@bose.res.in}},
{\bf {\normalsize Sunandan Gangopadhyay}$^{b,c}$\thanks{sunandan.gangopadhyay@gmail.com, sunandan@iiserkol.ac.in, sunandan@associates.iucaa.in}},
{\bf {\normalsize Pradip Mukherjee}$^{d}$\thanks{mukhpradip@gmail.com}}\\[0.2cm]
$^{a}$ {\normalsize S.N. Bose National Centre for Basic Sciences, 
Kolkata 700098, India}\\[0.2cm]
$^{b}$ {\normalsize Department of Physical Sciences,} \\
{\normalsize Indian Institute of Science Education and Research Kolkata,   }\\
{\normalsize Mohanpur 741246, Nadia, India}\\[0.2cm]
$^{c}${\normalsize Visiting Associate in Inter University Centre for Astronomy $\&$ Astrophysics (IUCAA),}\\
{\normalsize Pune, India}\\
$^{d}${\normalsize Department of Physics, Barasat Government College,}\\ 
{\normalsize 10, K. N. C. Road, Barasat, Kolkata 700124, India.}\\[0.3cm]
}


\date{}
\maketitle
\begin{abstract}
\noindent
A novel algorithm is provided to couple a Galilean invariant model with curved spatial background by taking nonrelativistic limit of a unique minimally coupled relativistic theory, which ensures
Galilean symmetry in the flat limit and canonical transformation of the original fields.  That the twin requirements are fulfilled is ensured by a new field, the existence of which was demonstrated recently from Galilean gauge theory.
The ambiguities and anomalies concerning the recovery of Galilean symmetry in the flat limit of spatial non relativistic  diffeomorphic theories, reported in the literature, are focused and resolved from a new angle.

\end{abstract}



\newpage

\section{Introduction}

Nonrelativistic diffeomorphism invariance (NRDI)\footnote{In this paper we shall be strictly confined to spatial NRDI.} has recently received considerable attention because of its myriad applications . These are as diverse as mesoscopic physics \cite{SW, F, HS,Can, GS, BMM1, BMM3,M1, M2, BM4,Wu}, nonrelativistic gravity \cite{C1,C2}, Newton-Cartan geometry \cite{N1,N2,N3,K,K1,K2,N4,N5,ABP,ABP1, BMM2,BM5}, Horava Lifshitz gravity \cite{horava, AHH, BM4}, to name a few. A very important sector is fractional quantum Hall effect (FQHE), where we require diffeomorphism invariance in space. Such theories provide a way of coupling nonrelativistic field theories to background space and the consequent effective field theory becomes a powerful tool for these applications.
Since Galilean symmetry is a subset of the bigger nonrelativistic diffeomorphism symmetry, it is not just desirable, but essential, that the flat limit of NRDI should yield Galilean invariance. But this appears to be riddled with ambiguities and problems in the literature \cite{SW,BM4,AHH,andreev2,B1}.

To appreciate this issue, we recall that, lacking a definite prescription, NRDI was originally introduced in an ad-hoc fashion in \cite{SW}.
An outcome of such an approach was the occurence of an unusual transformation for the space component of the vector field. It was unusual in the sense that, going to the flat limit, the standard Galilean result (under boosts) could not be reproduced. To overcome this situation, a particular relation between the gauge parameter and boost parameter was suggested. While this cancelled the anomalous boost term it simultaneously created another problem; namely, the original gauge freedom was lost. This is hardly surprising since gauge transformation and general coordinate (diffeomorphism) transformation are independent and any relation between them is bound to affect one or the other. Here the problem merely gets shifted from the boost sector to the gauge sector. Though several studies based on \cite{SW} have been reported, this inconsistency was never dealt with.

Nonrelativistic difeomorphism was introduced long back by Cartan \cite{C1, C2}
when he formulated Newtonian gravity as a geometric theory in what is now called Newton-Cartan (NC) spacetime. Still  the question of coupling a Galilean symmetric theory with curved spacetime is a difficult one. 
Since time in non relativistic theories has an absolute status, NC spacetime is absolutely foliated (since the direction of time flow is absolute), defining the Galilean frame. To construct spatial NRDI, we require to anticipate the Galilean frame. 
 Recently we have tackled this issue by developing a systematic algorithm in \cite{BMM1,BMM2,BMM3,BM4,BM5}. 
We gauged the Galilean symmetry of a dynamical model for the purpose. This Galilean gauge theory (GGT) when applied to the model of \cite{SW} gives a diffeomorphism invariant theory in  space \cite{BMM3,BM4} which is satisfactory in all respects. This has been amply demonstrated \cite{BMM3,BM4} including the points of departure from \cite{SW}. Since a large number of recent works \cite{HS,GS,M1,M2} in the theory of fractional Hall effect is based on \cite{SW}, these points of departure assume quite a lot of significance.

One of the most attractive features of Galilean gauge theory is the complete ease with which one can go to the flat limit. This is not accidental. We have shown that this theory is amenable to a geometric interpretation. The new fields introduced during localization can be identified with the elements of the NC geometry in the vielbein approach. 
 Moreover the coordinate system for the curved spacetime automatically emerges as the Galilean frame.

With this perspective in mind let us elaborate the distinctive feature of the present work. When we localize the symmetry of the action corresponding to the motion of a 2 + 1 dimensional trapped electron, the geometry dictates an extra term in the action depending linearly on a new field \cite{BMM1}. Such a field is non-existant in  usual approaches \cite{SW,HS,GS,M1,M2}. Apart from this new field which transforms anomalously, all the original fields have canonical transformation properties.  The occurrence of this new field is mandatory, otherwise NRDI can not be obtained consistently. This is strikingly demonstrated for the example of a free theory. Surprisingly, the analysis of NRDI for a free theory was never performed. This will be shown here in section 3 not from our algorithm of GGT \cite{BM4}, but rather by taking the nonrelativistic limit of a free relativistic theory.


This naturally brings us to the issue of interpreting models with NRDI as derived by taking the nonrelativistic limit of some relativistic model. Here also there is lack of any systematic methodology. Indeed different relativistic models leading to the same nonrelativistic diffeomorphism invariant model have been reported \cite{AHH,andreev2}.
The original motivation for studying nonrelativistic limits was to provide a semblance of justification for the ad-hoc and unusual construction of NRDI models. But, as shown here, this also fails the test of a correct flat limit. Other approaches to the problem are also plagued with the same problem \cite
{AHH,andreev2}. As we show, taking the nonrelativistic limit and the flat limit can
be done in a unified and consistent  manner in GGT, which is unique. 

The last point highlights a significant new result of our paper and needs elaboration. We have shown that several relativistic actions may be reduced to the same nonrelativistic NRDI action. However, there is only one relativistic action for which there is a consistent flat limit and the reduced `metric'  transforms canonically. Also, the reduced metric is unique and does not depend on whether the reduction has been carried out for a free theory or an interacting theory. This unique reduction process can be identified following the method presented here. 



In section 2 we state the problem providing necessary technical details. Section 3 analyses the free theory. NRDI in this case was not discussed earlier in the literature. Neither can it be obtained by simply setting the vector (gauge) field to zero in the action of \cite{SW}. Hence this is an important example. Already in this example the new field is introduced. The flat limit is examined and the relativistic origin of NRDI is also clarified.
The calculations are then extended to the interacting model in section 4. Both the flat limit and relativistic origin are discussed. Results are compared with the existing ones in the literature \cite{SW, AHH,andreev2}. An alternative action whose nonrelativistic limit also yields the same NRDI model is given. This is obtained by an ad-hoc process without the systematic basis that pervades this paper.
Expectedly, the flat limit poses problems and once again illustrates the vagaries of an ad-hoc procedure. 
In section 5 we show that the relativistic metric leading to NRDI with the correct flat limit admits an ADM decomposition. Finally, section 6 contains our conclusions.




\section{Statement of the problem}
Let us begin with a statement of the problem. A consistent way to couple nonrelativistic particles to an external gauge field and metric tensor
such that it manifests a nonrelativistic version of general coordinate invariance has, as already mentioned, generated considerable interest.
This invariance has deep consequences and nontrivial applications, particularly
in the context of unitary Fermi gas, FQHE and Newton-Cartan geometry.

There are, however, subtle traps and pitfalls in a consistent formulation
of nonrelativistic diffeomorphism invariance. There are two issues involved.
First, a smooth flat limit should exist that recovers the original Galilean symmetry of the nonrelativistic model. Secondly, the relativistic origin of
nonrelativistic diffeomorphism invariance should be clarified. On both these
counts the discussions in the existing literature \cite{SW,AHH,andreev2,B1} are rather controversial and dubious. These issues are highlighted by taking a specific example.

Consider a system of nonrelativistic particles coupled to an external gauge
field $A_\mu$ in flat space \cite{SW},
\begin{eqnarray}
S &=&  \int\!dt\,d\x\, \left[\frac i2 \psi^\+ \dlr_t\psi
  - A_0\psi^\+\psi 
  - \frac{1}{2m}(D_{i}\psi)^{\dagger}D_i\psi\right].
\label{r1} 
\end{eqnarray} 
where $\psi^\+\dlr_t\psi = \psi^\+\d_t\psi-\d_t\psi^\+\psi$
and $D_i\psi=\partial_{i}\psi+iA_{i}\psi$.

\noindent This action has a local gauge invariance
\begin{eqnarray}
\psi\rightarrow\psi'=e^{i\alpha}\psi~,~A_0 \rightarrow A_{0}'=A_{0}-\dot{\alpha}~,~A_i \rightarrow A_{i}'=A_{i}-\partial_{i}\alpha
\label{r2} 
\end{eqnarray} 
where the gauge parameter is space time dependent, $\alpha(x, t)$.

The above action also has a Galilean invariance where the coordinates transform as
\begin{eqnarray}
t\rightarrow t'&=& t-\varepsilon\nonumber\\
x^i \rightarrow x'^{i}&=&x^i +\varepsilon^{i}+{\lambda^{i}}_{j}x^{j}-v^{i}t
\label{r3} 
\end{eqnarray} 
under which the infinitesimal transformations of the fields are given by
\begin{eqnarray}
\delta\psi&=&\psi'(x, t)-\psi(x, t)=-\xi^{\mu}\partial_{\mu}\psi=
\varepsilon\dot{\psi}-(\eta^i -v^i t)\partial_{i}\psi \nonumber\\
\delta A_0&=& A'_0(x, t)-A_{0}(x, t)=\varepsilon\dot{A}_0-(\eta^i -v^i t)\partial_{i}A_{0}+v^{i}A_i \nonumber\\
\delta A_i&=& A'_i(x, t)-A_{i}(x, t)=\varepsilon\dot{A}_i-(\eta^j -v^j t)\partial_{j}A_{i}+{\lambda_{i}}^{j}x_{j}.
\label{r4} 
\end{eqnarray} 
Here $\xi^{0}=\varepsilon$, $\eta^i =\varepsilon^{i} +{\lambda^{i}}_{j}x^j$ and $\xi^i = \eta^i -v^i t$. The Galilean parameters corresponding to spatial translations ($\varepsilon^{i}$), rotations (${\lambda^{i}}_{j}$) and boosts ($v^i$) are global. This completes the standard description of Galilean symmetry in a nonrelativistic model.

It is now possible to consider a system analogous to eq.(\ref{r1}) but defined in a curved three dimensional manifold with the spatial line element
\begin{eqnarray}\label{line-element}
ds^2 = g_{ij}(t,\x)\, dx^i\, dx^j ~.
\end{eqnarray}
The action of the system is given by
\begin{eqnarray}
S &=&  \int\!dt\,d\x\, \sqrt{\tilde g}\left[\frac i2 \psi^\+ \dlr_t\psi
  - A_0\psi^\+\psi 
  - \frac{g^{ij}}{2m}(\d_i\psi^\+-iA_i\psi^\+)(\d^j\psi+iA^j\psi)\right]
\label{globalaction2a} 
\end{eqnarray} 
where the determinant of $g_{ij}$ is denoted as $\tilde{g}$ ($\tilde{g}=det g_{ij}$). This action is invariant under the following set of infinitesimal transformations \cite{SW}
\begin{subequations}\label{nonrel-gci-A}
\begin{align}
\delta\psi &= i\alpha\psi -\xi^k\d_k\psi 
\label{nonr-A}\\
\delta A_0 &= -\dot\alpha-\xi^k\d_k A_0 - A_k \dot\xi^k 
\label{nonrel-gci-A0-A}\\
\delta A_i &= -\d_i\alpha-\xi^k\d_k A_i - A_k\d_i\xi^k  +mg_{ik}\dot{\xi^k}
\label{nonrel-SW-A}\\
\delta g_{ij} &= -\xi^k\d_k g_{ij} -g_{ik}\d_j\xi^k -
 g_{kj}\d_i\xi^k .
 \label{m}
\end{align}
\end{subequations}
Here both the gauge parameter $\alpha$ and the diffeomorphism parameter $\xi^i$
(corresponding to the shift $x^i \rightarrow x^i +\xi^i$) are space time dependent. The above transformations (involving $\xi^i$) are referred as
nonrelativistic spatial diffeomorphisms. Now, for consistency, the Galilean set
(\ref{r4}) under which eq.(\ref{r1}) is invariant should be reproduced as a restricted class of eq.(\ref{nonrel-gci-A})
under which eq.(\ref{globalaction2a}) is invariant. The standard replacement of
$g_{ij}$ by the flat metric $\eta_{ij}$ followed by taking $\xi^i$ in
eq.(\ref{nonrel-gci-A}) as space time independent 
should reproduce eq.(s)(\ref{r1}) and (\ref{r4}) with $\varepsilon=0$
(since only spatial transformations are considered). Let us specifically
concentrate on the boosts, which is the only nontrivial sector, by choosing 
$\xi^{i}=-v^{i}t$ in eq.(\ref{nonrel-gci-A}). The passage from 
eq.(\ref{globalaction2a}) to eq.(\ref{r1}) is smooth when $g_{ij}\rightarrow \eta_{ij}$. However, the same cannot be said about eq.(\ref{nonrel-gci-A}) to eq.(\ref{r4}). Specifically,
eq.(\ref{nonrel-SW-A}) yields
\begin{eqnarray}
\delta A_i &= -\d_i\alpha + t v^k\d_k A_i -mv_{i}
\label{nonrel-SW-AB}
\end{eqnarray}
which clashes with eq.(\ref{r4}) due to the presence of the last term.

It has been suggested in \cite{SW} that by choosing $\alpha=-mv_{i}x^{i}$,
the extra piece in eq.(\ref{nonrel-SW-AB}) gets cancelled. However, this does
not solve the problem. If this is done then the original gauge symmetry
(\ref{r2}) of $A_i$ is lost. 
This shows that a consistent flat space limit of eq.(\ref{nonrel-gci-A}) does not exist.


To mitigate a lack of systematic development it was suggested in \cite{SW} that there was a relativistic origin of eq.(\ref{globalaction2a}). The idea was to consider a relativistic field theory of a free complex scalar field $\phi$ in an external
four dimensional metric $g_{\mu\nu}$. With $x^\mu=(ct,\,\x)$ and a mostly positive metric signature $({-}{+}{+}{+})$, the invariant length is given by
\begin{eqnarray}\label{element}
ds^2 = g_{\mu\nu}(t,\x)\, dx^{\mu}\, dx^{\nu} ~.
\end{eqnarray}
The relevant action is given by
\begin{equation}\label{relat-S}
  S = -\int\!d^4x\, \sqrt{-g^{(4)}}\, \left( g^{\mu\nu}\d_\mu\phi^*\d_\nu\phi 
       +m^2c^2\phi^*\phi\right)
\end{equation}
which is invariant under the infinitesimal general coordinate transformations
\begin{eqnarray}\label{relat-gci}
  \delta\phi = - \xi^\lambda\d_\lambda \phi
\end{eqnarray}
\begin{eqnarray}\label{relat-gcin}
    \delta g_{\mu\nu} = -\xi^\lambda\d_\lambda g_{\mu\nu} 
    -g_{\lambda\nu}\d_\mu\xi^\lambda - g_{\mu\lambda}\d_\nu\xi^\lambda.
\end{eqnarray}
Making a change of variables 
\begin{equation}\label{Psipsi}
  \phi = e^{-imcx^0}\frac\psi{\sqrt{2mc}} 
  = e^{-imc^2t}\frac\psi{\sqrt{2mc}}\,
\end{equation}
and choosing a metric
\begin{equation}\label{rmetr}
  g_{\mu\nu} = \left( \begin{array}{ccc}
   -1 -\dfrac{2A_0}{mc^2} 
     +O(c^{-4})&
   & -\dfrac{A_i}{mc}+ O(c^{-3})\rule[-6mm]{0mm}{6mm}\\
   -\dfrac{A_i}{mc}+ O(c^{-3}) & & 
   g_{ij} + O(c^{-2})
  \end{array}\right)
\end{equation}
it is possible to reproduce action (\ref{globalaction2a}) in the 
$c\to\infty$ limit. Also, the transformations (\ref{nonrel-gci-A}) are obtained from eq.(s)(\ref{relat-gci})-(\ref{Psipsi}) provided the following identification is used
\begin{equation}\label{used}
\xi^{\mu}=\left(\dfrac{\alpha}{mc}, \xi^{i}\right). 
\end{equation}
One may therefore interpret eq.(s)(\ref{globalaction2a}) 
and (\ref{nonrel-gci-A}) as suitable nonrelativistic limits of 
eq.(s)(\ref{relat-S}), (\ref{relat-gci}) and (\ref{relat-gcin}) .

This line of reasoning also fails to provide a satisfactory resolution of the
flat limit problem. To obtain this limit, we replace $g_{\mu\nu}\rightarrow\eta_{\mu\nu}$ in eq.(\ref{relat-S}) which corresponds to setting $A_{\mu}=0$ and $g_{ij}=\delta_{ij}$ in eq.(\ref{rmetr}). 
This ensures that the action (\ref{relat-S}) has a viable flat limit
which yields the usual theory of complex scalars. The corresponding situation in
eq.(\ref{globalaction2a}) is, however, problematic. Taking $A_{\mu}=0$ trivialises the flat limit of the action (\ref{globalaction2a}) and fails to reproduce the expected result (\ref{r1}). This failure is hardly surprising since the lack of a proper flat limit is an intrinsic defect and cannot be bypassed by merely algebraic gymnastics.

\section{Free Theory}

The problem of formulating a theory having NRDI has been clearly posed - how to consistently take the flat limit and recover the standard Galilean symmetry? A related issue is to understand the
NRDI as a nonrelativistic limit of the diffeomorphism invariance of some relativistic theory. As will be shown, these are connected isuues.

The shortcomings and pitfalls highlighted in the previous section are a consequence of the ad-hoc approach to 
NRDI. The first thing, therefore, is to develop a systematic algorithm for 
NRDI. This was earlier presented in a series of papers involving two of the present authors \cite{BMM1,BMM3,BM4,BMM2}. We shall now use this algorithm to present a resolution of the problems.

Let us first recall that NRDI was initially discussed in the context of the model (\ref{globalaction2a}) to eventually analyse the trapping of electrons
on a two dimensional plane that is the forerunner of the FQHE problem. A simpler
theory would be to consider the noninteracting or free theory. Unfortunately
this cannot be obtained by just putting $A_0=A_i=0$ 
in eq.(s)(\ref{globalaction2a}) and (\ref{nonrel-gci-A}). It is easy to check that NRDI does not hold. This manifests the seriousness of the issue. While one may get away with the construction (\ref{globalaction2a}) and its symmetry (\ref{nonrel-gci-A}) (forgetting the flat limit), it is not possible to construct a nonrelativistic diffeomorphism invariant free theory in this manner.

We now apply the formalism \cite{BMM1,BMM3,BM4,BMM2} to the free theory. First, the free nonrelativistic theory on flat background is written
\begin{eqnarray}
S &=&  \int\!dt\,d\x\, \left[\frac i2 \psi^\+ \dlr_t\psi
   - \frac{1}{2m}\d_i\psi^\+ \d_i\psi\right]
\label{globalaction2} 
\end{eqnarray} 
which is invariant under the Galilean transformations (\ref{r3})
and (\ref{r4}).

The next step is to gauge the galilean symmetry. This is done by taking the parameters of Galilean transformations to be functions of space time.
Naturally, the original invariance under (global) Galilean transformations
is destroyed, since the transformation of the derivatives gets modified. The symmetry is recovered by replacing the ordinary derivatives by suitable covariant derivatives. This entails introduction of new fields. Also, the measure gets altered due to a nontrivial Jacobian. The transformation of the new fields are determined to ensure the (local) invariance of the theory. Interestingly, the fields entering the Jacobian transform exactly as the square root of the determinant of a spatial metric (\ref{m}). This leads to a geometrical interpretation of the gauged theory. Taking into account all considerations, we finally obtain \cite{BMM3,BM4}
\begin{eqnarray}\label{free-L}
  S &=&  \int\!dt\,d\x\, \sqrt{\tilde g}\left[\frac i2 \psi^\+ \dlr_t\psi
  - B_0\psi^\+\psi 
  - \frac{g^{ij}}{2m}(\d_i\psi^\+-iB_i\psi^\+)(\d_j\psi+iB_j\psi)\right]\nonumber\\
&& + \int\!dt\,d\x\, \sqrt{\tilde g}~ \frac{i}{2}~\Delta^{k} 
 \left[\psi^{\dagger}(\partial_{k}\psi +iB_k \psi) - (\partial_{k}\psi^{\dagger}-iB_k \psi^{\dagger})\psi\right].
\end{eqnarray}
The $B$ and $\Delta$ variables are new (external) fields introduced by the gauging process. Setting them to zero leads to a flat metric (as we shall see in the subsequent discussion) and reproduces the flat space theory (\ref{globalaction2}). The above action is invariant       
under the following infinitesimal  transformations \cite{BMM3,BM4}
\begin{subequations}\label{nonrel-gci}
\begin{align}
\delta\psi &= -\xi^k\d_k\psi 
\label{nonr}\\
\delta B_0 &=-\xi^k\d_k B_0 - B_k \dot\xi^k 
\label{nonrel-gci-A0}\\
\delta B_i &= -\xi^k\d_k B_i - B_k\d_i\xi^k 
\label{nonrel}\\
\delta \Delta_i &=\dot{\xi_i} -\xi^{k}\partial_{k}\Delta_{i} -\Delta_{k}\partial_{i}\xi^{k}
\label{nonrel-gci-Ai}
\end{align}
\end{subequations}
together with the transformation for $g_{ij}$ (\ref{m}).
Taking the flat limit, as discussed above, immediately reproduces the Galilean theory (\ref{globalaction2}) invariant under the transformation (\ref{r4}).


\subsection{Relativistic origin of nonrelativistic diffeomorphism invariance }

We now show the obtention of eq.(\ref{free-L}) by taking an appropriate nonrelativistic limit of a relativistic action. Once again, contrary to earlier studies \cite{SW,AHH,andreev2}, our approach will be guided by a systematic algorithm. This is now elaborated. 

Before considering the abstraction of eq.(\ref{free-L}), we take the simpler and familiar theory (\ref{globalaction2}). Indeed this free theory may be derived from its relativistic version
\begin{equation}\label{relat-S-fla}
  S = -\int\!d^4x\,  \left( \eta^{\mu\nu}\d_\mu\phi^*\d_\nu\phi 
       +m^2c^2\phi^*\phi\right)
\end{equation}
by making the change of variables (\ref{Psipsi}) and finally taking the 
$c\to\infty$ limit.

It is now expected that, by taking the curved space generalization of 
eq.(\ref{relat-S-fla}), the action (\ref{free-L}) should be derivable.
Thus, we begin from the action (\ref{relat-S})
which is invariant under the general coordinate transformations
\begin{equation}\label{gct1}
x^{\mu}\rightarrow x^{\mu}+\xi^{\mu}(x, t) 
\end{equation}
where the field $\phi$ and the metric $g_{\mu\nu}$ transform as 
eq.(\ref{relat-gci}).

Once again making the change of variables (\ref{Psipsi}) and taking the 
$c\to\infty$ limit (keeping $g_{\mu\nu}$ fixed), yields
\begin{eqnarray}\label{free-L1}
S = \int\!dt\,d\x\, \sqrt{\tilde g}\left[\frac i2 \psi^\+ \dlr_t\psi
-\frac{mc^2 (g^{00}+1)}{2}\psi^\+\psi -\frac{i}{2}c g^{0i}[\psi^{\dagger}\partial_{i}\psi-(\partial_{i}\psi^{\dagger})\psi]
  -\frac{g^{ij}}{2m}\d_i\psi^\+\d_j\psi\right].
\end{eqnarray}
Requiring that the above equation reproduces eq.(\ref{free-L}) leads to the following relations 
\begin{eqnarray}\label{sol}
g^{00}&=&-1+\frac{2}{mc^2}\left[B_0 +\frac{B^{i}B_{i}}{2m}+\Delta^{i}B_{i}\right]\\
g^{0i}&=&-\frac{1}{c}\left[\frac{B^i}{m}+\Delta^i\right].
\end{eqnarray}
Hence the inverse metric is given by
\begin{equation}\label{Ginv}
  g^{\mu\nu} = \left( \begin{array}{ccc}
   -1 + \dfrac{2B_0}{mc^2} 
     +\dfrac{B^i B_i}{m^2c^2}+\dfrac{2\Delta^{i}B_{i}}{mc^2}+O(c^{-4})&
   & -\left[\dfrac{B^i}{mc}+\dfrac{\Delta^i}{c}\right]+ O(c^{-3})\rule[-6mm]{0mm}{6mm}\\
   -\left[\dfrac{B^i}{mc}+\dfrac{\Delta^i}{c}\right] + O(c^{-3}) & & 
   g^{ij} + O(c^{-2})
  \end{array}\right)
\end{equation}
where $B^i\equiv g^{ij}B_j$. The metric therefore reads
\begin{equation}\label{Gmetr}
  g_{\mu\nu} = \left( \begin{array}{ccc}
   -1 -\dfrac{2B_0}{mc^2} 
     +\dfrac{\Delta^{i}\Delta_{i}}{c^2}+O(c^{-4})&
   & -\left[\dfrac{B_i}{mc}+\dfrac{\Delta_i}{c}\right]+ O(c^{-3})\rule[-6mm]{0mm}{6mm}\\
   -\left[\dfrac{B_i}{mc}+\dfrac{\Delta_i}{c}\right] + O(c^{-3}) & & 
   g_{ij} + O(c^{-2})
  \end{array}\right)
\end{equation}
where $\Delta_i = g_{ij}\Delta^j$.

With this form of the relativistic metric, the theory for the relativistic
complex scalar field given by the action (\ref{relat-S}) reduces to the action (\ref{free-L}) in the nonrelativistic limit $c\to\infty$.

\noindent We now proceed to take the nonrelativistic limit of the relativistic infinitesimal transformations (\ref{relat-gci}). 
Taking $\xi^\mu$ of the form
\begin{equation}
\xi^\mu = \left( 0\,,\ \xi^i \right)\label{gct}
\end{equation}
it is easy to see that the infinitesimal transformation for the relativistic field $\phi$ (\ref{relat-gci}) yields the infinitesimal transformation for the nonrelativistic field $\psi$ (\ref{nonr}).
Now taking the $0i$ component of the metric in eq.(\ref{relat-gcin})
and keeping terms of the $\mathcal{O}(c^{-1})$ on both sides yields 
the infinitesimal transformation for the fields $B_i$ and  $\Delta_i$ (\ref{nonrel}, \ref{nonrel-gci-Ai}).
Taking the $00$ component of the metric in eq.(\ref{relat-gcin})
and keeping terms of the $\mathcal{O}(c^{-2})$ on both sides yields 
the infinitesimal transformation for the field $B_0$
(\ref{nonrel-gci-A0}) and the following equation for the 
infinitesimal transformation of $\Delta^i$
\begin{equation}
(\delta \Delta^i)\Delta_i + \Delta^i \delta \Delta_i
=\Delta_i [\dot{\xi^i} -\xi^k \partial_{k}\Delta^i  - \Delta^k \partial_{k}\xi^i].
\label{Del1}
\end{equation}
Substituting the form for the infinitesimal transformation of
$\Delta_i$ (\ref{nonrel-gci-Ai}) in the above equation leads to
\begin{equation}
\delta \Delta^i
=\dot{\xi^i} -\xi^k \partial_{k}\Delta^i  + \Delta^k \partial_{k}\xi^i .
\label{Del2}
\end{equation}
The same result also follows on using $\Delta_j =g_{ji}\Delta^{i}$ and computing the $\delta$-variation on both sides
\begin{equation}
\delta \Delta^i
=g^{ij}\delta\Delta_{j}-g^{ij}\delta g_{jk}\Delta^{k}.
\label{Del2al}
\end{equation}
Thus one can understand the set of infinitesimal transformations (\ref{nonr})-(\ref{m}) as a nonrelativistic limit of the relativistic general coordinate invariance. 

The last point is to demonstrate the consistency of the flat limit.
Previously it was mentioned that this limit is accomplished by setting to zero the new fields $B$ and $\Delta$, besides taking $g_{ij}\rightarrow\delta_{ij}$. That the action (\ref{free-L})
and the transformations (\ref{nonrel-gci}) correctly pass to the flat
expressions (\ref{globalaction2}) and (\ref{r4}) was already seen. The new point is that this can be complemented with the flat limit taken here. The structure of the metric (\ref{Gmetr}) is such that it indeed reduces to the flat metric $\eta_{\mu\nu}$ when $B$ and $\Delta$ are set to zero. This is a nontrivial point in discussing flat limits from a nonrelativistic reduction since the metric $g_{\mu\nu}$ is now endowed with a specific form 
(like eq.(\ref{Gmetr})). We mentioned it earlier and will return to it later.
Now the theory (\ref{relat-S}) passes over to eq.(\ref{relat-S-fla}) whose nonrelativistic version is eq.(\ref{globalaction2}), the flat limit of eq.(\ref{free-L}). Everything falls into place like a jigsaw puzzle.

Let us recapitulate the sequence of arguments that yields a closed chain. This is best expressed diagrammatically:
\begin{figure}[H]
  \centering
  \includegraphics[width=15cm]{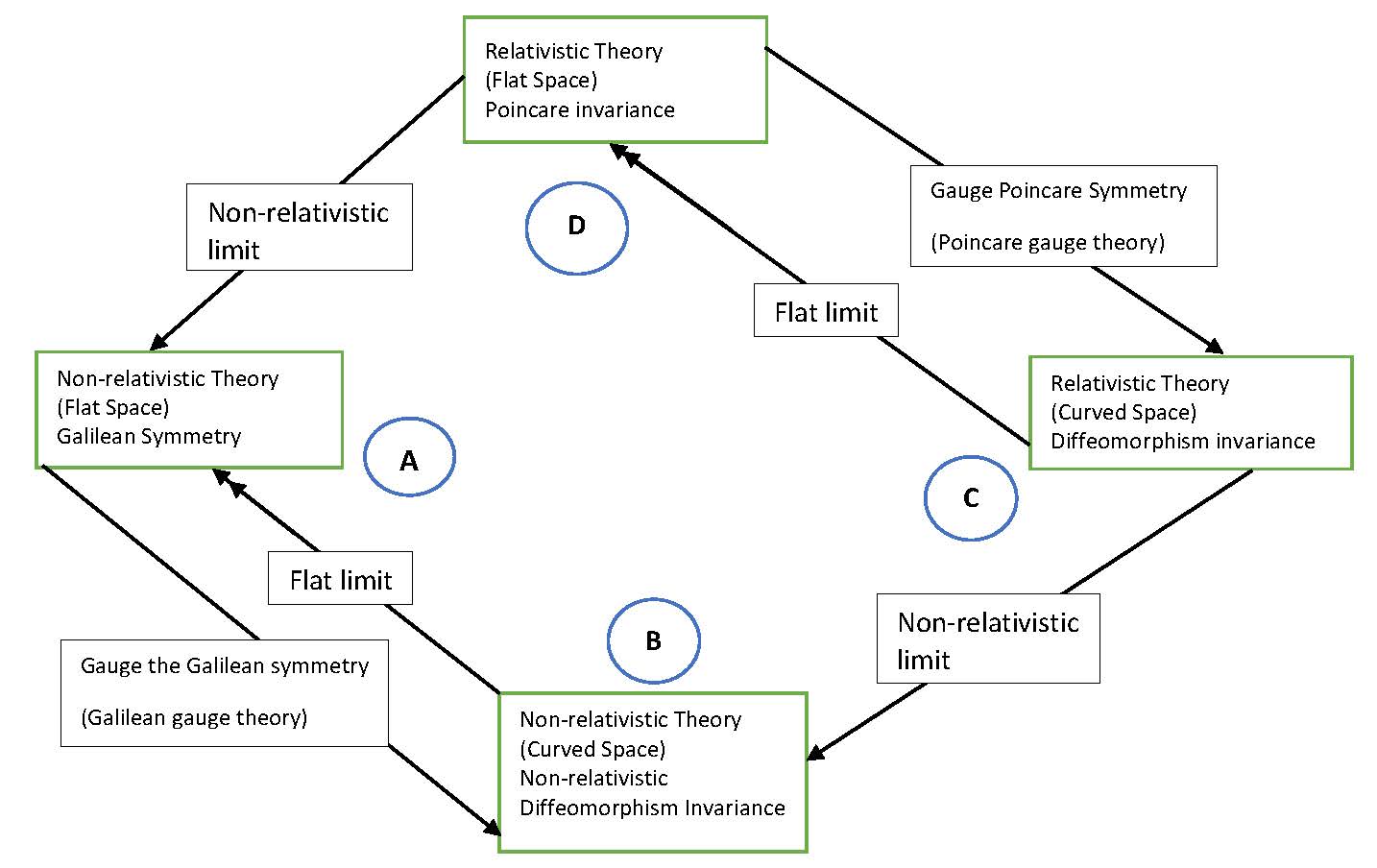}
  \caption{Interplay between different symmetries.}
  \label{fig1}
\end{figure}

We started from the box (A) to reach (B) by following the single headed arrow. To obtain (C) whose NR limit would yield (B), we use our knowledge of (A) to abstract (D). Knowing (D) it is straightforward to find (C) by following the double headed arrow.




\section{Interacting Theory}


The important example of particles interacting with a gauge field is now taken up. The analysis will follow exactly the same route
as for the free theory succintly depicted in the diagram. Following the same algorithm we first consider the flat space (Euclidean) action already defined in eq.(\ref{r1}). This action
is invariant under global Galilean transformations (\ref{r4}) as can be checked explicitly \cite{BMM3}. Note that the theory is also invariant under $U(1)$ gauge transformation (\ref{r2}). Naturally these two types of symmetry are mutually exclusive, the former is a symmetry under space time transformations while the latter corresponds to phase rotation in the internal space.

Once again by localising the Galilean symmetry of the action and geometrical reinterpretation we can reformulate the theory in Newton-Cartan manifold on the unique spatial foliation corresponding to the Galilean frame 
with the spatial line element (\ref{line-element}). 
The action of the system invariant under diffeomorphism of the manifold is given by \cite{BMM3,BM4}
\begin{eqnarray}\label{free-L-A}
  S   &=&  \int\!dt\,d\x\, \sqrt{\tilde g}\left[\frac i2 \psi^\+ \dlr_t\psi
  - (A_0 +B_0)\psi^\+\psi 
  - \frac{g^{ij}}{2m}[\d_i\psi^\+-i(A_i +B_i)\psi^\+][\d_j\psi+i(A_j +B_j)\psi]\right]\nonumber\\
&& + \int\!dt\,d\x\, \sqrt{\tilde g}~ \frac{i}{2}~\Delta^{k} 
 \left[\psi^{\dagger}[\partial_{k}\psi +i(A_k +B_k) \psi] - [\partial_{k}\psi^{\dagger}-i(A_k +B_k) \psi^{\dagger}]\psi\right].
\end{eqnarray}

\noindent The action (\ref{free-L-A}) is invariant under the following infinitesimal  transformation
\begin{eqnarray}
\delta A_i &= -\d_i\alpha-\xi^k\d_k A_i - A_k\d_i\xi^k 
\label{nonrel-gci-A-r}
\end{eqnarray}
together with the infinitesimal transformations 
for $\psi$, $A_0$, $B_0$, $B_i$, $\Delta_i$ and $g_{ij}$
given in eq.(s)(\ref{nonr-A}, \ref{nonrel-gci-A0-A}, \ref{nonrel-gci-A0}, \ref{nonrel}, \ref{nonrel-gci-Ai}, \ref{m}). 

Observe that the fields $A$ and $B$ appear in the combination $(A+B)$.
However we do not rename it since $A$ and $B$ have distinct transformation properties. They have different roles when the flat limit is taken.

The above transformation rule (\ref{nonrel-gci-A-r}) for the field $A_i$ 
is the usual transformation for the vector potential under gauge transformation together with diffeomorphism and differs 
from that in eq.(\ref{nonrel-SW-A}). 
There the absence of the new field $\Delta^i$ required an anomalous transformation law  for the field $A_i$. However, no such modification in the transformation laws is required once the field $\Delta^i$ is present which was introduced by the gauging prescription.

Taking the flat limit poses no problems. Set the new fields $B_0$, $B_i$
and $\Delta_i$ to zero and replace $g_{ij}\rightarrow\delta_{ij}$ which immediately yields eq.(\ref{r1}).
Also, there is no anomalous transformation as happened earlier 
(since eq.(\ref{nonrel-SW-A}) is now replaced by eq.(\ref{nonrel-gci-A-r})).
Results for the free theory are reproduced by putting the external gauge
field $A_{\mu}$ to zero. This completes the passage from box B to the box A (see Figure 1).

\subsection{Relativistic origin}
To reinterpret the action (\ref{free-L-A}) as a nonrelativistic limit of some relativistic theory, the earlier procedure is adopted. We initially construct the box D (see Figure 1). This is given by
\begin{equation}\label{relat-S-flat}
S = -\int\!d^4x\, \left[\eta^{\mu\nu}(D_\mu\phi)^* D^\nu\phi 
+m^2c^2\phi^*\phi\right]
\end{equation}
where 
\begin{eqnarray}
D_\mu\phi=\partial_\mu\phi +i\mathcal{A}_{\mu}\phi~.
\label{cov-qq}
\end{eqnarray}
This action has a $U(1)$ gauge invariance given by eq.(\ref{r2}) 
(replace $\psi$ by $\phi$).
The passage from box D to box A is now discussed. 
Substituting the form of the field $\phi$ in terms of the nonrelativistic field $\psi$ by eq.(\ref{Psipsi}) and making the identification
\begin{eqnarray}
\mathcal{A}_0=\dfrac{A_0}{c}~,~ \mathcal{A}_i =A_i
\label{ident-F}
\end{eqnarray}
yields 
\begin{eqnarray}\label{flatlimit}
S &=& -\int\!d^3x~dt\, 
\left[-\frac i2 \psi^\+ \dlr_t\psi 
+\frac{1}{2m}\partial_{i}\psi^{*}\partial_{i}\psi 
+\frac{1}{2m}\left(iA^{i}(\partial_i \psi^{*}\psi -\psi^{*}\partial_i \psi)+ \left(-\dfrac{{A_{0}}^2}{c^2}+ {A_{i}}^2\right)\psi^{*}\psi\right)\right.\nonumber\\
&&\left.-\frac{i}{c}A_0 \left[\psi^{*}(ic\psi -\frac{1}{c}\dot{\psi})+
\left(ic\psi^{*}+\frac{1}{c}\dot{\psi}^{*}\right)\psi\right]\right].
\end{eqnarray}
Taking the $c\to\infty$ limit, the  action (\ref{r1})
is obtained, thereby completing the passage from box D to box A.

It is now straightforward to construct the box C and study its limit to box B.
The appropriate theory pertaining to box C is given by first lifting 
eq.(\ref{relat-S-flat}) from a flat to a curved background
\begin{equation}\label{relat-S-qq}
S = -\int\!d^4x\, \sqrt{-g}\, \left[ g^{\mu\nu}(D_\mu\phi)^*D_\nu\phi 
+m^2c^2\phi^*\phi\right].
\end{equation}
Apart from the $U(1)$ gauge invariance this action is also invariant under general coordinate transformations (\ref{gct1}) with eq.(s)(\ref{relat-gci}),
(\ref{relat-gcin}) and
\begin{equation}\label{newtr}
\delta A_{\mu}=-\xi^{\nu}\partial_{\nu}A_{\mu}-A_{\nu}\partial_{\mu}\xi^{\nu}.
\end{equation}
Now taking eq.(s)(\ref{Psipsi}, \ref{ident-F}) and the form of the metric 
(\ref{Ginv}), (\ref{Gmetr}) it is possible to reproduce the action 
(\ref{free-L-A}) in the $c\to\infty$ limit.

It is important to note that the structure of the metric (\ref{Ginv}), (\ref{Gmetr}) that effects the nonrelativistic passage is the same for the free theory and the interacting theory. This is indicative of the geometrical nature
of the NRDI. Exactly the same analysis (from eq.(\ref{Del1}) 
till eq.(\ref{Del2al}))
gets repeated to correctly reproduce the transformations of $B_0$, $B_i$,
$\Delta_i$ and $g_{ij}$. The transformations of 
$A_0$ eq.(\ref{nonrel-gci-A0-A}), $A_i$ 
eq.(\ref{nonrel-gci-A-r}) and $\psi$ eq.(\ref{nonr-A}) are trivially obtained from eq.(\ref{relat-gci}) and eq.(\ref{newtr}), recalling eq.(\ref{Del1}).

As happened for the free theory, the flat limit is consistently implemented since the metric is unchanged. Setting the new fields ($B_0$, $B_i$ and $\Delta_i$) to
zero also ensures transition of eq.(s)(\ref{Ginv}, \ref{Gmetr}) to their flat versions. The action (\ref{relat-S-qq}) reduces to eq.(\ref{relat-S-flat}) whose nonrelativistic limit was seen to be eq.(\ref{r1}). Thus the complete chain indicated by the box diagram is completed.

\subsection{Relativistic origin: an alternative action}
In the above we have shown how the spatially diffeomorphic nonrelativistic theories can be obtained on reduction of appropriate relativistic theories. However the simple fact that this may be done does not gurantee the authenticity of the theory, the transformation rules of the metric tensor must also be reduced to the appropriate nonrelativistic transformations. It is possible to discuss an alternative route for the nonrelativistic reduction. But we see that the transformation rules are anomalous for the alternative reduction. 

From our previous analysis, it is possible to identify an alternative path. Start from the action (\ref{relat-S}) with the metric
\begin{equation}\label{Gmetr-A}
  g_{\mu\nu} = \left( \begin{array}{ccc}
   -1 -\dfrac{2(A_0 + B_0)}{mc^2} 
     +\dfrac{\Delta^{i}(A_i + B_{i})}{c^2}+O(c^{-4})&
   & -\left[\dfrac{(A_i +B_i)}{mc}+\dfrac{\Delta_i}{c}\right]+ O(c^{-3})\rule[-6mm]{0mm}{6mm}\\
   -\left[\dfrac{(A_i +B_i)}{mc}+\dfrac{\Delta_i}{c}\right] + O(c^{-3}) & & 
   g_{ij} + O(c^{-2})
  \end{array}\right)
\end{equation} 
and its inverse 
\begin{equation}\label{Ginv-A}
  g^{\mu\nu} = \left( \begin{array}{ccc}
   -1 + \dfrac{2C_0 }{mc^2} 
    +\dfrac{C^i C_i }{m^2c^2}+\dfrac{2\Delta^{i}C_i}{mc^2}+O(c^{-4})&
   & -\left[\dfrac{C^i}{mc}+\dfrac{\Delta^i}{c}\right]+ O(c^{-3})\rule[-6mm]{0mm}{6mm}\\
   -\left[\dfrac{C^i}{mc}+\dfrac{\Delta^i}{c}\right] + O(c^{-3}) & & 
  g^{ij} + O(c^{-2})
  \end{array}\right)
  \end{equation}
where $C_0 =A_0 +B_0$, $C_i =A_i +B_i$ and $C^i\equiv g^{ij}C_j$.
In the limit $c\to\infty$ the action (\ref{relat-S}) 
reproduces eq.(\ref{free-L-A}).

\noindent We now proceed to take the nonrelativistic limit of the relativistic infinitesimal transformations (\ref{relat-gci}). 
Taking $\xi^\mu$ of the form in eq.(\ref{used})
where $\alpha$ is fixed in the limit  $c\to\infty$, it is easy to see that the infinitesimal transformation for the relativistic field $\phi$ 
(\ref{relat-gci}) yields the infinitesimal transformation for the nonrelativistic field $\psi$ (\ref{nonr-A}).
Now taking the $0i$ component of the metric in eq.(\ref{relat-gcin})
and keeping terms of the $\mathcal{O}(c^{-1})$ on both sides yields 
the infinitesimal transformation for the fields $A_i$, $B_i$ and  $\Delta_i$ ((\ref{nonrel-gci-A-r}), \ref{nonrel}, \ref{nonrel-gci-Ai}).
Taking the $00$ component of the metric in eq.(\ref{relat-gcin})
and keeping terms of the $\mathcal{O}(c^{-2})$ on both sides yields 
the infinitesimal transformation for the fields $A_0$ and $B_0$
given in eq.(s)(\ref{nonrel-gci-A0-A}, \ref{nonrel-gci-A0}). 

Expectedly, the flat limit cannot be consistently implemented. As pointed out earlier, the flat limit corresponds to setting the newly introduced fields
($B_0$, $B_i$, $\Delta_i$) to zero and putting the curved spacetime metric $g_{\mu\nu}=\eta_{\mu\nu}$. As seen from eq.(\ref{Gmetr-A}, \ref{Ginv-A}), the two are not compatible.


\section{Connection of the relativistic metric with ADM decomposition}

It is interesting to note that the metric (and its inverse)
found during the nonrelativistic reduction from the relativistic actions
admit an ADM-decomposition . Let us recall that the ADM construction
is given by
\begin{equation}
ds^2 =-N^2 c^2 dt^2 +g_{ij}(dx^i +N^i c dt)(dx^j +N^j c dt)
\label{ADM1}
\end{equation}
where $N$ and $N^{i}$ are the lapse and shift variables. In this case the
metric and its inverse are given by  
\begin{equation}\label{ADMinv}
  g_{\mu\nu} = \left( \begin{array}{ccc}
   -N^2 +g_{ij}N^{i}N^{j}&
   & g_{ij}N^j\rule[-6mm]{0mm}{6mm}\\
   g_{ij}N^j & & 
   g_{ij} 
  \end{array}\right)
\end{equation}
and 
\begin{equation}\label{ADM}
  g^{\mu\nu} = \left( \begin{array}{ccc}
   -\dfrac{1}{N^2}&
   & \dfrac{N^i}{N^2}\rule[-6mm]{0mm}{6mm}\\
   \dfrac{N^i}{N^2} & & 
   \left(g_{ij}-\dfrac{N^i N^j}{N^2}\right) 
  \end{array}\right).
\end{equation}
Comparing with the expressions (\ref{Ginv}, \ref{Gmetr}) for   
$g^{\mu\nu}$ and $g_{\mu\nu}$, we find
\begin{eqnarray}\label{ident1}
N&=& \sqrt{1+\frac{2B_0}{mc^2}+\frac{B^i B_i}{m^2 c^2}+\frac{2\Delta^i B_i}{mc^2}} +\mathcal{O}(c^{-4})
\end{eqnarray}  
\begin{eqnarray}\label{ident}
N_i &=& -\frac{B_i}{mc}-\frac{\Delta_i}{c}+\mathcal{O}(c^{-3}).
\label{ident2}
\end{eqnarray}   


\noindent Computing the variations of the above equations, we get
\begin{eqnarray}\label{var}
\delta N=-\xi^{j}\partial_{j}N
\end{eqnarray}  
\begin{eqnarray}\label{ident-A}
\delta N_i = -\partial_{i}\xi^{j}N_{j} -\xi^{j}\partial_{j}N_{i}
-\dfrac{\dot{\xi}^{j}}{c} g_{ij}
\label{var-A}
\end{eqnarray}
which is in agreement with the variations of the 
ADM variables \cite{horava}\footnote{To make a comparison with \cite{horava} one has to put $\xi^{0}=0$ there which is the case examined here.}.


  
\section{Conclusion} 
In this work we have elaborated on the flat space limit of nonrelativistic diffeomorphism invariance (NRDI) which is expected to yield the Galilean invariance. Usual constructions have not provided a detailed analysis on this aspect which is fundamental to a consistent formulation of NRDI. The transformation rule for boosts becomes anomalous for which there is no satisfactory explanation. Linking the boost parameter with the gauge parameter,
as suggested in \cite{SW}, cannot save the day because then the original gauge symmetry is lost. An alternative attempt based on interpreting NRDI as a limit of nonrelativistic invariance also fails. As we have discussed, the origin of these failures essentially lies in the ad-hoc formulation of obtaining NRDI. 

In a set of papers involving two of us \cite{BMM1,BMM3,BM4,BMM2,BM5}, a systematic algorithm to discuss NRDI was developed by gauging the Galilean symmetry. The galilean gauge theory formulation of NRDI required the introduction of new fields. The transformations of the new fields is spelled out from the approach itself. This, together with the transformations of the original fields, ensures NRDI. This theory has a smooth flat limit. The new fields are set to zero in this limit and the diffeomorphism invariant theory goes over to the Galilean invariant theory where the fields have usual transformation properties.

In this paper we provide a consistent interpretation of NRDI from a relativistic origin. The appropriate relativistic theory was obtained systematically by following the flow chart at the end of section 3.
We started with the complex  Klein-Gordon field minimally coupled to external gravity. By devising a foliation of the relativistic space time we expand the metric components in appropriate powers of $c^{-1}$. The metric components are obtained by term by term comparision 
with the nonrelativistic model in question, and extracting the rest energy in the dominant phase factor, complete equivalence 
with the nonrelativistic model along with its invariances is shown. From the reduction procedure we also show the physical necessity of the new field emerging from the Galilean gauge theory algorithm in yet another way.
Finally we have shown that in the reduction process we actually perform an ADM decomposition of the metric, which conforms to the nonrelativistic fixed foliation of spacetime \cite{horava}.
This justifies our identification of the spatial slice in taking the nonrelativistic limit. It is important to point out that here also, taking the flat limit has no problems. The new fields introduced in the Galilean gauge theory enter naturally in the relativistic metric as a correction to the flat metric. The important point, however, is that this correction does not involve the original fields, as happens in the other approaches\cite{SW,AHH,andreev2}. Consequently, the flat limit is consistently implemented. The new fields have to be set equal to zero which automatically ensure the passage of the curved metric to the flat metric. If the metric involved the original fields, obviously the flat metric cannot be reproduced.

The genesis of the pitfalls and/or shortcomings of earlier approaches is clearly illuminated by the free theory which was analysed here in considerable details. We may note in passing that NRDI of a free model was never discussed in the literature because the focus was on a nonrelativistic charged particle under an external electromagnetic field in curved space which proved to be useful in discussing fractional quantum Hall effect as well as 
formulating the Newton-Cartan geometry. 
Indeed, as shown here, if we simply set the external gauge field to zero, we do not recover NRDI in a free theory. The introduction of new fields is mandatory to achieve this invariance. These fields naturally emerge in the reduction process from the relativistic to the nonrelativistic case. Also, in the present approach, NRDI for the free case is obtained by simply setting the external gauge field to zero. This is a very desirable feature.
Finally, the fact that the reduced 
 metric (\ref{Ginv},\ref{Gmetr}, in such a discussion, remains unchanged irrespective of the theory being free or interacting, highlights the geometrical nature of NRDI.



\bigskip

\section*{ Acknowledgments}
S.G. acknowledges the support by DST SERB under Start Up Research Grant (Young Scientist), File No.YSS/2014/000180. The authors thank the referee for very useful comments.

\end{document}